\documentclass[10pt]{iopart}
\expandafter\let\csname equation*\endcsname=\relax
\expandafter\let\csname endequation*\endcsname=\relax
\usepackage{amsmath,amssymb}

\begin{document}

\title{Metric symmetry by design in general relativity}

\author{Viktor T. Toth$^1$\footnote{Corresponding author ({\tt vttoth$@$vttoth.com})}}

\address{$^1$Ottawa, Ontario K1N 9H5, Canada}

\date{\today}

\begin{abstract}
The usual derivation of Einstein's field equations from the Einstein--Hilbert action is performed by silently assuming the metric tensor's symmetric character. If this symmetry is not assumed, the result is a new theory, such as Einstein's attempted Unified Field Theory or Moffat's Nonsymmetric Gravitational Theory. Explicitly enforcing the constraint by means of a Lagrange-multiplier term restores Einstein's field equations, but the multiplier appears as an additional, unconstrained antisymmetric term. We briefly discuss the possible significance of this term with respect to a nonvanishing cosmological angular momentum, a sourced spin current, the nonsymmetric nature of the Einstein pseudotensor characterizing the energy-momentum of the gravitational field, and possible implications on attempts to obtain a quantum theory of gravity.
\end{abstract}



It is widely known that the field equations of gravitation, that is, Einstein's field equations, can be derived using the variational principle from the Einstein--Hilbert Lagrangian \cite{einstein1915,Hilbert1915} and the corresponding action,
\begin{align}
{\cal S}_{\rm grav}=\int d^4x {\cal L}_{\rm grav}=\int d^4x \sqrt{-g}\frac{1}{2\kappa}(R-2\Lambda),
\end{align}
where $g$ is the determinant of the metric $g_{\mu\nu}$, $R$ is the Ricci-tensor, and $\Lambda$ is the cosmological constant, with $\kappa=8\pi G$.

The derivation is conceptually straightforward, though there are technical subtleties, arising, in particular, due to the fact that the Ricci scalar $R$ depends on both first and second derivatives of the metric tensor. Generally, the presence of such higher derivatives is bad news, as they can lead to an unstable theory with a Hamiltonian unbounded from below (Ostrogradsky instability \cite{ostrogradsky1850}). General relativity is an exception, as derivatives of the metric tensor beyond the second derivative cancel out in the resulting field equations. (This can also be inferred from the fact that general relativity can be re-expressed in terms of the connection coefficients in the Palatini formalism \cite{Palatini1919}, and in this case, only first derivatives are present in the Lagrangian.) The Euler-Lagrange equations that correspond to ${\cal S}_{\rm grav}$ are given, as usual \cite{LR1989}, by functional differentiation with respect to the metric tensor and its derivatives:
\begin{align}
\frac{1}{\sqrt{-g}}&\left[\frac{\partial{\cal L}_{\rm grav}}{\partial g_{\mu\nu}}-\frac{\partial}{\partial x^\kappa}\frac{\partial{\cal L}_{\rm grav}}{\partial g_{\mu\nu,\kappa}}+\frac{\partial^2}{\partial x^\kappa\partial x^\lambda}\frac{\partial{\cal L}_{\rm grav}}{\partial g_{\mu\nu,\kappa\lambda}}\right]
+\text{[matter terms]}=0.
\end{align}
The matter terms in the general relativistic action are usually given in the form of the matter stress-energy tensor, defined as $T^{\mu\nu}=-2(-g)^{-1/2}\delta {\cal S}_{\rm matter}/\delta g_{\mu\nu}$, where ${\cal S}_{\rm matter}$ is the action integral representing the matter content. The gravitational part, in turn, after a somewhat tedious but straightforward derivation, yields the well-known Einstein tensor, ultimately leading to Einstein's field equations for gravitation:
\begin{align}
R^{\mu\nu}-\tfrac{1}{2}Rg^{\mu\nu}+\Lambda g^{\mu\nu}=8\pi GT^{\mu\nu}.
\end{align}

There is, however, a delicate issue here that remains largely unexplored in the general relativity literature, starting with Hilbert's publication all the way back in 1915, the first to invoke the action principle (albeit without a cosmological constant) to derive the field equations of gravitation \cite{Hilbert1915}.

We ``know'' that the metric tensor is symmetric\footnote{In Hilbert's own words: ``{\em die zehn zuerst von Einstein eingef{\"u}lhrten Gravitation-potentiale $g_{\mu\nu}$ ($\mu,\nu = 1, 2, 3, 4$) mit symmetrischen Tensorcharakter}'', i.e., the ten gravitational potentials first introduced by Einstein, $g_{\mu\nu}$ ($\mu,\nu=1,2,3,4$), with symmetric tensor character. Similarly, Rosenfeld \cite{Rosenfeld1940} explored the properties, including the symmetry properties, of the energy-momentum tensor, but the symmetry of the metric tensor was simply assumed.}: that is to say, $g_{\mu\nu}=g_{\nu\mu}$, which appears to follow from the definition of the invariant line element, $ds^2=g_{\mu\nu}dx^\mu dx^\nu=g_{(\mu\nu)}dx^\mu dx^\nu$, since $g_{[\mu\nu]}dx^\mu dx^\nu\equiv 0$ identically\footnote{Note that we follow the usual convention for symmetrizing indices, $g_{(\mu\nu)}=\tfrac{1}{2}(g_{\mu\nu}+g_{\nu\mu})$ and $g_{[\mu\nu]}=\tfrac{1}{2}(g_{\mu\nu}-g_{\nu\mu})$.}.

However, just because we assume that $g_{\mu\nu}$ is symmetric does not make it so. The dynamical theory of gravitation does not automatically imply its geometric interpretation\footnote{Indeed, Einstein himself argued that geometrization is not essential: ``{\em Sie haben vollst{\"a}ndig recht. Es ist verkehrt zu glauben, dass die `Geometrisierung' etwas Wesentliches bedeutet. Es ist nur eine Art Eselsbr{\"u}cke zur Auffindung numerischer Gesetze. Ob man mit einer Theorie 'geometrische' Vorstellungen verbindet, ist [unleserlich] Privatsache.}'', or ``You are perfectly right. It is wrong to think that the `geometrization' has significant meaning. It is only a kind of a clue helping us find numerical laws. Whether you connect a 'geometric' view to a theory is entirely a private matter.''  -- Einstein's letter to Rosenbach, April 8, 1926, as published (with translation) by Lehmkuhl \cite{lehmkuhl2014}.}. And if $g_{\mu\nu}$ is ``only'' the tensor field mediating the gravitational interaction, not the metric of spacetime, why would it be symmetrical? Specifically, when we apply the variational principle to a functional that depends on $g_{\mu\nu}$ and its first and second derivatives, the sixteen components of $g_{\mu\nu}$ vary independently unless they are constrained. Variation of the Einstein-Hilbert term with respect to the unconstrained metric tensor yields additional contributions to the field equations, leading, e.g., to Einstein's attempt to develop a unified field theory \cite{AMT1966} or to Moffat's nonsymmetric generalization of Einstein's theory of gravitation \cite{Moffat1995}.

On the other hand, the metric tensor can be explicitly constrained to be symmetrical by way of introducing a Lagrange-multiplier into the Lagrangian formulation of the theory.

The use of Lagrange-multiplier terms in general relativity to enforce constraints, of course, has been explored in the literature over the more than 100 years since Einstein and Hilbert published their respective papers. In particular, Lanczos \cite{Lanczos1949} used Lagrange-multipliers to recast Einstein's theory using a third-order field tensor that is antisymmetric in two of its indices. Later, Kichenassamy \cite{Kichenassamy1986}, relying in part on the work of Lanczos, systematically explored the use of Lagrange-multipliers, establishing a consistent relationship between varying the metric components vs. varying the connection coefficients.

Yet surprisingly, none of these authors appear to have addressed this more basic matter of the assumed symmetry of the metric tensor.

This attitude is perhaps justifiable if the connection coefficients and the terms constructed from them, notably the Riemann curvature tensor and the associated Ricci-tensor and Ricci-scalar, are formed in such a manner that the metric tensor is effectively symmetrized, though that still allows for possible nonsymmetric contributions through the integration measure $\sqrt{-g}$ that is an essential part of the generally covariant action.

A more direct alternative, however, is to make the assumption that $g_{[\mu\nu]}=0$ explicit by way of introducing an appropriate Lagrange-multiplier term in the Einstein-Hilbert Lagrangian, namely a tensor-valued term $\lambda_{\mu\nu}$:
\begin{align}
{\cal S}_{\rm grav}=\frac{1}{2\kappa}\int d^4x \sqrt{-g}(R-2\Lambda+\lambda^{\mu\nu}g_{[\mu\nu]}).
\end{align}
Let us explore the consequences of this proposal.

Variation with respect to the Lagrange-multiplier's nondynamical degrees of freedom trivially yields the expected constraint,
\begin{align}
g_{[\mu\nu]}=0,\label{eq:gsym}
\end{align}
now explicitly restricting the metric to be symmetrical. Enforcing this constraint after computing the variation of the Einstein--Hilbert term, we find that any additional contributions vanish, and we are left with the usual Einstein field tensor. (This can also be verified through tedious but straightforward computer algebra calculations.) That still leaves open an intriguing question: What is the variation of the newly introduced Lagrange-multiplier term with respect to $g_{\mu\nu}$, and how does that contribute to the theory's field equations?

The answer can be obtained by calculating the following functional derivative:
\begin{align}
\frac{\delta}{\delta g_{\alpha\beta}}(\sqrt{-g}\lambda^{\mu\nu}g_{[\mu\nu]})&{}=
\left[\frac{\delta}{\delta g_{\alpha\beta}}(\sqrt{-g}\lambda^{\mu\nu})\right]g_{[\mu\nu]}
+\sqrt{-g}\lambda^{\mu\nu}\frac{\delta}{\delta g_{\alpha\beta}}g_{[\mu\nu]}.
\end{align}
The first term in this expression can be discarded due to (\ref{eq:gsym}). As to the second term, it reads
\begin{align}
\sqrt{-g}\lambda^{\mu\nu}\frac{\delta}{\delta g_{\alpha\beta}}g_{[\mu\nu]}
=\sqrt{-g}\lambda^{\mu\nu}\tfrac{1}{2}(\delta^\alpha_\mu\delta^\beta_\nu-\delta^\alpha_\nu\delta^\beta_\mu)=\sqrt{-g}\lambda^{[\alpha\beta]}.
\end{align}
Shouldn't this term, then, appear in the Einstein field equations?

It is tempting to dismiss the term simply by declaring the Lagrange-multiplier $\lambda_{\mu\nu}$ to be symmetric in the first place: $\lambda_{\mu\nu}=\lambda_{(\mu\nu)}$, hence $\lambda_{[\mu\nu]}=0$. But in this case, we lose the constraint altogether, since we'd have $\lambda^{\mu\nu}g_{[\mu\nu]}=\lambda^{(\mu\nu)}g_{[\mu\nu]}=0$ identically, without constraining $g_{\mu\nu}$ to be symmetric. So $\lambda_{\mu\nu}$ must remain a generic, nonsymmetric tensor.

But that implies that Einstein's field equations must be modified:
\begin{align}
R_{\mu\nu}-\tfrac{1}{2}Rg_{\mu\nu}+\Lambda g_{\mu\nu}+\lambda_{[\mu\nu]}=8\pi G T_{\mu\nu}.\label{eq:EFE}
\end{align}
What is the meaning of $\lambda_{[\mu\nu]}$, a nondynamical antisymmetric term?

As the name implies, a constraint term like this typically constrains the system's equations of motion; e.g., we can use a Lagrange multiplier to constrain the Lagrangian of the mass in a rigid pendulum, confining its motion. In our case, however, the $\lambda_{[\mu\nu]}$ term in the field equations appears to play a different role.

First, we note that all the other terms are symmetric and conserved: The Einstein tensor $G^{\mu\nu}=R^{\mu\nu}-\tfrac{1}{2}Rg^{\mu\nu}$ is divergence free, $G^{\mu\nu}_{;\mu}=0$ (as usual, the semicolon indicates covariant differentiation), as is the cosmological constant term since $g^{\mu\nu}_{;\mu}=0$.

In other words, the left-hand side of Eq.~(\ref{eq:EFE}) is the sum of a conserved, symmetric term $G_{\mu\nu}+\Lambda g_{\mu\nu}$, and an unconstrained antisymmetric term in the form $\lambda_{[\mu\nu]}$. Correspondingly, we expect the right-hand side to have a similar structure. This implies conservation of the symmetrized matter stress-energy-momentum tensor:
\begin{align}
T^{(\mu\nu)}_{;\mu}=0.
\end{align}

We cannot make the same statement about the antisymmetric part, $T_{[\mu\nu]}$, however. If it is present, it is entirely unconstrained by the gravitational field equations, since it is ``eaten'' by $\lambda_{[\mu\nu]}$ and $\lambda_{\mu\nu}$ is an unconstrained nondynamical term. Thus we end up with two distinct equations:
\begin{align}
R_{\mu\nu}-\tfrac{1}{2}Rg_{\mu\nu}+\Lambda g_{\mu\nu} &{}= 8\pi G T_{(\mu\nu)},\\
\lambda_{[\mu\nu]} &{}= 8\pi GT_{[\mu\nu]}.
\end{align}

This leads to the striking result that not only is the matter stress-energy-momentum tensor not constrained to be symmetrical by the gravitational field equations, but that {\em Einstein's gravitational field is unaffected by the antisymmetric part of a generalized stress-energy-momentum tensor.} This antisymmetric part represents a net rotation, net angular momentum, sometimes called the spin tensor, $S_{\mu\nu}=T_{[\mu\nu]}.$

Historically, the presence of a net rotation was often accommodated by introducing the concept of torsion, a term arising from dropping the symmetry constraint not on the metric tensor but on the connection coefficients. This approach can be formally introduced through the Palatini formulation of general relativity, leading to Einstein-Cartan theory \cite{Cartan1922,Cartan1923,Cartan1925} when the connection is not assumed to be symmetrical. When the connection is assumed to be symmetrical, the Palatini formalism also yields Einstein's field equations but again, there is a catch. In a rigorous treatment, it is not sufficient to express only in words the constraint that restricts the connection coefficients to the (symmetric) Christoffel-symbols. Rather, it must be incorporated into the Lagrangian using a Lagrange-multiplier \cite{Safko1976,Burton1998}.


To further explore the meaning of the antisymmetric stress-energy-momentum tensor, let us consider the angular momentum current:
\begin{align}
J^{\lambda\mu\nu}=x^\mu T^{\lambda\nu}-x^\nu T^{\lambda\mu}.
\end{align}
Conservation of this current can be expressed as
\begin{align}
\nabla_\lambda J^{\lambda\mu\nu} = 0.
\end{align}
If $T^{\mu\nu}$ is conserved, i.e., if $\nabla_\mu T^{\mu\nu}=0$, the divergence of the angular momentum current becomes
\begin{align}
\nabla_\lambda J^{\lambda\mu\nu} = 2T^{[\mu\nu]},
\end{align}
thus conservation of $J^{\lambda\mu\nu}$ implies the symmetry $T^{[\mu\nu]} = 0$.

This condition, however, is not necessarily satisfied. In particular it is violated when a spin current $S^{\lambda\mu\nu}$ with sources is present:
\begin{align}
J^{\lambda\mu\nu}=(x^\mu T^{\lambda\nu}-x^\nu T^{\lambda\mu}) + S^{\lambda\mu\nu}.
\end{align}
Since $\nabla_\lambda S^{\lambda\mu\nu}\ne 0$ in the presence of sources, $\nabla_\lambda J^{\lambda\mu\nu}=0$ implies a nonzero $T^{[\mu\nu]}$.

This may be addressed by the somewhat {\em ad hoc} construction of the symmetric, conserved Belinfante--Rosenfeld stress-energy-momentum tensor \cite{Rosenfeld1940,Belinfante1940}, which may be written as
\begin{align}
\Theta^{\mu\nu}=T^{\mu\nu}+\tfrac{1}{2}\nabla_\lambda (S^{\mu \nu \lambda }+S^{\nu \mu \lambda }-S^{\lambda \nu \mu }).
\end{align}

This construction becomes unnecessary in our formalism. As we have shown, formally incorporating the symmetry of the metric tensor into the gravitational field Lagrangian ensures that the gravitational field depends only on $T^{(\mu\nu)}$; the nonsymmetric contribution to the stress-energy-momentum tensor remains unconstrained by the {\em gravitational} field equations. This of course does not mean that $T^{[\mu\nu]}$ is unconstrained or nondynamical: it simply means that any constraint on $T^{[\mu\nu]}$ must come from the theory of matter, not from the theory of gravity. Insofar as gravity is concerned, any nonsymmetric contribution due, e.g., to the presence of a spin current is therefore easily accommodated.


Our assertion, therefore, is that in the standard, Einstein--Hilbert form of the general relativistic action, even if we take it for granted that gravitation is fundamentally geometric in nature, it 
%
%
is not sufficient to simply state in words that the metric tensor is symmetrical. This constraint, if desired, must be explicitly incorporated into the Lagrangian. This leads us to an intriguing conclusion. If we do introduce such a constraint, we end up with gravitational field equations that govern only the symmetric part of the stress-energy-momentum tensor; the antisymmetric part, representing rotation, remains unconstrained by gravity. If, in contrast, we choose not to modify the action but no longer silently assume that the metric is symmetric or torsion is absent, we end up with a new theory. Depending on specific details, the result may be Einstein's attempt to create a classical unified field theory; Moffat's nonsymmetric gravitational theory; Einstein-Cartan theory; or some other theoretical framework that accounts for the new independent degrees of freedom and ascribes to them physical meaning.

The one thing that we are not allowed to do if we favor a rigorous treatment is to silently assume that $g_{[\mu\nu]}=0$ without accounting for this constraint in the Lagrangian. This approach would be mathematically inconsistent.

Finally, we note that the presence of nonsymmetric terms may open interesting new directions of investigation. One possibility is that it may grant new legitimacy to the (nonsymmetric) Einstein pseudotensor that, in some contexts, is used to represent the energy-content of the gravitational field. The presence of the constraint terms or the presence of additional degrees of freedom if the constraint terms are omitted, may also have an impact on attempts to develop viable quantum theories of gravitation.

\section*{Acknowledgements}

VTT acknowledges the generous support of  	
Vladimir Andreev, Plamen Vasilev and other Patreon patrons.

\section*{References}
\bibliography{refs}

\begin{thebibliography}{10}

\bibitem{einstein1915}
Albert {Einstein}.
\newblock {Die Feldgleichungen der Gravitation}.
\newblock {\em Sitzungsberichte der K{\"o}niglich Preussischen Akademie der
  Wissenschaften}, pages 844--847, December 1915.

\bibitem{Hilbert1915}
David {Hilbert}.
\newblock {Die Grundlagen der Physik}.
\newblock {\em {Nachrichten von der Gesellschaft der Wissenschaften zu
  G{\"o}ttingen Mathematisch-Physikalische Klasse}}, 3:395--407, 1915.

\bibitem{ostrogradsky1850}
M.~{Ostrogradsky}.
\newblock {M{\'e}moire sure les {\'e}quations diff{\`e}rentielles relatives au
  probl{\'e}me des isop{\'e}rim{\`e}tres}.
\newblock {\em {M{\'e}moires de l'acad{\'e}mie imp{\'e}riale des sciences de
  Saint-P{\'e}tersbourg}}, {Sixi{\'e}me s{\'e}rie Sciences math{\'e}matiques,
  physiques et naturelles, Tome VI}({Premi{\`e}re partie: sciences
  math{\'e}matiques et physiques, Tome IV}):385--517, 1850.

\bibitem{Palatini1919}
Attilio {Palatini}.
\newblock {Deduzione invariantiva delle equazioni gravitazionali dal principio
  di Hamilton}.
\newblock {\em {Rendiconti del Circolo Matematico di Palermo (1884-1940)}},
  43(1):203--212, December 1919.

\bibitem{LR1989}
David Lovelock and Hanno Rund.
\newblock {\em {Tensors, Differential Forms, and Variational Principles}}.
\newblock Dover Publications, 1989.

\bibitem{Rosenfeld1940}
L{\'e}on {Rosenfeld}.
\newblock {Sur le tenseur d'impulsion--{\'e}nergie}.
\newblock {\em {M{\'e}moires Acad. Roy. De Belgique}}, 18:1--30, 1940.
\newblock English translation by D. H. Delphenich.

\bibitem{lehmkuhl2014}
Dennis {Lehmkuhl}.
\newblock {Why Einstein did not believe that general relativity geometrizes
  gravity}.
\newblock {\em Studies in the History and Philosophy of Modern Physics},
  46:316--326, May 2014.

\bibitem{AMT1966}
M.~A. {Tonnelat}.
\newblock {\em {Einstein's Unified Field Theory}}.
\newblock Gordon and Breach Science Publishers, 1966.

\bibitem{Moffat1995}
J.~W. {Moffat}.
\newblock {A new nonsymmetric gravitational theory}.
\newblock {\em Physics Letters B}, 355:447--452, February 1995.

\bibitem{Lanczos1949}
Cornelius {Lanczos}.
\newblock {Lagrangian Multiplier and Riemannian Spaces}.
\newblock {\em Reviews of Modern Physics}, 21(3):497--502, July 1949.

\bibitem{Kichenassamy1986}
S.~{Kichenassamy}.
\newblock {Lagrange multipliers in theories of gravitation}.
\newblock {\em Annals of Physics}, 168(2):404--424, May 1986.

\bibitem{Cartan1922}
Elie Cartan.
\newblock Sur une g{\'e}n{\'e}ralisation de la notion de courbure de riemann et
  les espaces {\`a} torsion.
\newblock {\em Comptes rendus hebdomadaires des s{\'e}ances de l'Acad{\'e}mie
  des sciences}, 174:593--595, 1922.
\newblock S{\'e}ance du 27 f{\'e}vrier 1922.

\bibitem{Cartan1923}
Elie Cartan.
\newblock Sur les vari{\'e}t{\'e}s {\`a} connexion affine et la th{\'e}orie de
  la relativit{\'e} g{\'e}n{\'e}ralis{\'e}e (premi{\`e}re partie).
\newblock {\em Annales scientifiques de l'{\'E}cole Normale Sup{\'e}rieure},
  40:325--412, 1923.

\bibitem{Cartan1925}
Elie Cartan.
\newblock Sur les vari{\'e}t{\'e}s {\`a} connexion affine, et la th{\'e}orie de
  la relativit{\'e} g{\'e}n{\'e}ralis{\'e}e (deuxi{\`e}me partie).
\newblock {\em Annales scientifiques de l'{\'E}cole Normale Sup{\'e}rieure},
  42:17--88, 1925.

\bibitem{Safko1976}
J.~L. {Safko} and F.~{Elston}.
\newblock {Lagrange multipliers and gravitational theory}.
\newblock {\em Journal of Mathematical Physics}, 17(8):1531--1537, August 1976.

\bibitem{Burton1998}
Howard~Steven Burton.
\newblock {\em {On The Palatini Variation and Connection Theories of Gravity}}.
\newblock PhD thesis, University of Waterloo, Waterloo, Ontario, Canada, 1998.
\newblock Also available through Library and Archives Canada.

\bibitem{Belinfante1940}
F.~J. {Belinfante}.
\newblock {On the current and the density of the electric charge, the energy,
  the linear momentum and the angular momentum of arbitrary fields}.
\newblock {\em Physica}, 7(5):449--474, May 1940.

\end{thebibliography}
\bibliographystyle{unsrt}

\end{document}